\documentclass[10pt,a4paper]{article}

\usepackage[margin=2.5cm]{geometry}
\usepackage{times}
\usepackage{graphicx}
\usepackage{amsmath,amssymb}
\usepackage{authblk}
\usepackage[numbers,sort&compress]{natbib}
\usepackage{hyperref}
\usepackage{listings}
\usepackage{xcolor}
\usepackage{enumitem}
\usepackage{pdflscape}
\usepackage{longtable}
\usepackage{booktabs} %
\usepackage{adjustbox}
\usepackage{subcaption}

\setlist[itemize]{noitemsep, topsep=2pt}
\setlist[enumerate]{noitemsep, topsep=2pt}

\lstdefinestyle{python}{
  language=Python,
  basicstyle=\ttfamily\footnotesize,
  keywordstyle=\color{blue},
  stringstyle=\color{red!60!black},
  commentstyle=\color{gray},
  breaklines=true,
  frame=single,
  showstringspaces=false
}
\title{\bfseries Kosmulator: A Python framework for cosmological inference with MCMC}

\author[1, 2]{Renier T. Hough}
\author[1]{Robert Rugg}
\author[1, 2]{Shambel Sahlu}
\author[1, 2]{Amare Abebe}
\affil[1]{\small Centre for Space Research, North-West University, Potchefstroom, South Africa, 2520}
\affil[2]{\small National Institute for Theoretical and Computational Sciences (NITheCS), Potchefstroom, South Africa, 2520}
\affil[ ]{\small E-mail: renierht@gmail.com}

\date{} % no date printed

\begin{document}

\maketitle

\begin{abstract}
We present \texttt{Kosmulator}, a modular and vectorised Python framework designed to accelerate the statistical testing of cosmological models. As the theoretical landscape expands beyond standard $\Lambda$CDM, implementing new expansion histories into traditional Einstein--Boltzmann solvers becomes a significant computational bottleneck. \texttt{Kosmulator} addresses this by leveraging array-native execution and efficient ensemble slice sampling (via \texttt{Zeus}) to perform rapid Bayesian inference. We validate the framework against the industry-standard \texttt{Cobaya} code using a combination of Type Ia Supernovae, Cosmic Chronometers, and Baryon Acoustic Oscillation (BAO) data. Our results demonstrate that \texttt{Kosmulator} reproduces \texttt{Cobaya}'s posterior constraints to within $\leq0.3\sigma$ statistical agreement on $H_{0}$ and $\Omega_{m}$ and $<0.6\%$ precision on $\chi^{2}$, while achieving a $\sim 4.5\times$ reduction in wall-clock time on a single CPU core compared to a standard MPI-parallelised baseline. Furthermore, we showcase the framework's utility by constraining the implicit power-law $f(Q)$ ``$f_1$CDM'' model and demonstrating its automated model selection capabilities (AIC/BIC). \texttt{Kosmulator} is introduced as a ``scientific sieve'' for rapid hypothesis testing, allowing researchers to efficiently filter theoretical candidates before deploying high-precision resources.
\end{abstract}

% ---------------------------------------------------------
\section{Introduction}

The standard $\Lambda$CDM model of cosmology has been remarkably successful in explaining a wide range of observations, from the Cosmic Microwave Background (CMB) to the large-scale structure of the Universe. However, the emergence of statistical tensions, most notably the Hubble tension ($H_0$) and the growth of structure tension ($S_8$), has spurred a resurgence of interest in alternative cosmological models and modified theories of gravity~\cite{Planck2020, Wong2020, Asgari2021, Valentino2021, Abbott2022, Abdalla2022, Riess2022, Abdul2025}.

Theoretical physicists are currently exploring a vast landscape of potential solutions, ranging from early dark energy~\cite{Poulin2019} and interacting dark sectors~\cite{Westhuizen2025a, Westhuizen2025b} to modifications of General Relativity such as $f(R)$~\cite{Hu2007}, $f(Q)$~\cite{Beltran2017} and $f(T)$~\cite{Bengochea2009} gravity. However, a significant computational bottleneck exists: testing these theoretical models against observational data typically requires their implementation into full Einstein--Boltzmann solvers such as \texttt{CLASS}\footnote{\texttt{CLASS} available at \url{https://github.com/lesgourg/class_public.git}}~\cite{Lesgourgues2011} or \texttt{CAMB}\footnote{\texttt{CAMB} available at \url{https://camb.readthedocs.io/en/latest/}}~\cite{Lewis2011}, to be utilized by frameworks like \texttt{MontePython}\footnote{\texttt{MontePython} available at \url{https://github.com/brinckmann/montepython_public}}~\cite{Brinckmann2019} and \texttt{Cobaya}\footnote{\texttt{Cobaya} available at \url{https://cobaya.readthedocs.io/en/latest/index.html}}~\cite{Torrado2021}, which are the industry standards for precision inference. These frameworks possess steep learning curves and adapting them for non-standard models often requires modifications in C and Fortran, which can be computationally expensive and time-consuming for preliminary model screening.

This technical barrier often slows the ``hypothesis-to-test'' cycle. Researchers may spend months implementing a model only to find it is ruled out by fundamental background observables. Therefore, there is a need for a lightweight, modular, and rapid inference framework that operates primarily at the background and phenomenological level.

In this paper, we introduce the core philosophy and modular architecture of \texttt{Kosmulator}\footnote{\texttt{Kosmulator} available at \url{https://github.com/renierht/Kosmulator}}, while providing a proof-of-concept demonstration of its utility for rapid model prototyping. Serving as a ``scientific sieve,'' the framework is designed to allow researchers to quickly test non-standard Friedmann equations and growth parameterizations using vectorised likelihoods before committing to more computationally intensive pipelines. We outline the fundamental usage of the package and present a baseline consistency check against the established \texttt{Cobaya} framework~\cite{Torrado2021} to demonstrate the reliability of our likelihood implementation for standard cosmologies. While a rigorous performance benchmark suite and high-dimensional parameter analysis are reserved for a forthcoming publication, this work establishes the framework's baseline viability for the community.

% ---------------------------------------------------------
\section{\texttt{Kosmulator}}

In this section, we describe the technical architecture of the \texttt{Kosmulator} MCMC package. Unlike standard Einstein--Boltzmann pipelines, which prioritize full perturbation-level accuracy at the cost of implementation complexity, \texttt{Kosmulator} is designed as a streamlined, pure-Python framework. Its primary objective is the rapid numerical evaluation of expansion histories and growth rates, even for models where the Friedmann equations are not analytically solvable.

The package is built on a modular architecture, enabling the direct comparison of theoretical predictions against late-time observational probes with minimal implementation effort. Building upon the foundational logic described in \cite{Hough2020}, this current version introduces a highly flexible sampling layer and streamlined dataset integration to facilitate broader community use.

\texttt{Kosmulator} utilizes both the \texttt{EMCEE}\footnote{\texttt{EMCEE} available at  \url{https://emcee.readthedocs.io/en/stable/}.}~\cite{Foreman2013} and \texttt{Zeus}\footnote{\texttt{Zeus} available at  \url{https://zeus-mcmc.readthedocs.io/en/latest/index.html}.}~\cite{Karamanis2021} samplers as interchangeable sampling engines. By providing access to both the Affine Invariant Ensemble Sampler (\texttt{EMCEE}) and the Ensemble Slice Sampler (\texttt{Zeus}), users can select the algorithm best suited to the specific topology of their model's parameter space. The latter, in particular, often provides superior performance for the highly non-linear or degenerate posteriors frequently encountered in modified gravity. For detailed discussions on the underlying methodologies, we refer the reader to \cite{Foreman2013, Karamanis2021}.

\subsection{Features}
\texttt{Kosmulator} contains the standard features found in competing software, such as:
\begin{itemize}
  \item \textbf{Multiple MCMC Backends} --- Support for both \texttt{Zeus} and \texttt{EMCEE}, allowing for automatic or user-controlled sampler selection.
  \item \textbf{Chain Reuse and Resume Capability} --- Reload, resume, or extend existing MCMC chains for reproducibility and efficient experimentation.
  \item \textbf{Publication-Ready Statistical Outputs and Visualisation} --- Automatic generation of corner plots, best-fit comparisons, autocorrelation diagnostics, and LaTeX-ready tables. Kosmulator's plotting pipeline utilises the \texttt{GetDist} package for post-processing and visualisation.
  \item \textbf{Automatic Parameter Injection and Dataset Awareness} --- Parameters (including nuisance parameters) are automatically added, fixed, or conditioned based on the selected observational datasets.
  \item \textbf{Support for a Wide Range of Observational Data}:
  \begin{itemize}
    \item Type Ia Supernovae (JLA, Pantheon, Pantheon+, Union3, DES-Y5)~\cite{Hicken2009, Neill2009, Conley2011, Scolnic2018, Brout2022, Rubin2025, DESY52024}
    \item Baryon Acoustic Oscillations (BAO)~\cite{Adame2025}
    \item DESI (Data Release 1 and 2)~\cite{Lodha2025, Abdul2025}
    \item Cosmic Chronometers (CC) / Observational Hubble Distance (OHD)~\cite{Moresco2020, Qi2023, Loubser2025a, Loubser2025b, Wang2026, Sharov2018}
    \item Cosmic Microwave Background (CMB; Planck 2018 low-$\ell$, high-$\ell$ TT, combined TTTEEE, and CMB lensing when used jointly with other datasets via \texttt{CLIK}\footnote{\texttt{CLIK} available at\url{http://pla.esac.esa.int/pla/}.}/\texttt{CLASS})~\cite{Lesgourgues2011, Planck2020}
    \item Big Bang Nucleosynthesis ($D/H$, BBN PryMordial, and AlterBBN\footnote{\texttt{AlterBBN} available at \url{https://github.com/espensem/AlterBBN.git}.}-based likelihoods)~\cite{Cooke2014, Abdul2025}
  \end{itemize}
\end{itemize}

While these functionalities are shared with other inference frameworks, such as \texttt{Cobaya}, which offers a broad suite of engines including nested samplers, and \texttt{MontePython}, which is tightly coupled with the \texttt{CLASS} code, \texttt{Kosmulator} distinguishes itself by matching these capabilities through a more streamlined, automated parameter-handling architecture.

\subsection{New features}
\subsubsection{Non-analytically solvable Friedmann equations} \label{Sec: User models}
Due to \texttt{Kosmulator}'s fully modular model architecture, any cosmological model that can be expressed as a standard \texttt{Python} function can be numerically solved and fitted to observational data. This design is particularly useful for cosmological models with implicit Friedmann equations. While inference frameworks such as \texttt{Cobaya} and \texttt{MontePython} can in principle be extended to support such models through modifications of external Einstein--Boltzmann solvers, doing so typically requires non-trivial development effort. \texttt{Kosmulator}, by contrast, was explicitly developed with numerical background-level model flexibility in mind.

In \texttt{Kosmulator}, adding a new cosmological model is as simple as defining a function of the form:
\begin{lstlisting}[style=python]
def LCDM_MODEL(z: Number, p: Dict[str, float]) -> Number:
    """Flat ΛCDM: E(z) = sqrt(Ωm(1+z)^3 + 1-Ωm)."""
    z = np.asarray(z, dtype=float)
    Om = float(p["Omega_m"])
    E2 = Om * (1.0 + z)**3 + (1.0 - Om)
    return np.where(E2 > 0, np.sqrt(E2), np.nan)
\end{lstlisting}
and registering it within the \texttt{User\_defined\_modules.py} script. Although this example corresponds to the standard $\Lambda$CDM model (already included in \texttt{Kosmulator}), the same structure applies to any background expansion history $E(z)$. 

For models lacking closed-form expansion histories, users can simply include an appropriate numerical solver within the model function. For example, in \cite{Sahlu2024} — the first scientific application of the open-sourced code — we considered a Power-law \(f(Q)\) model~\cite{Sahlu2025} defined implicitly by :
\begin{equation}\label{eq:f1CDM_model}
  E(z)^2 = \Omega_{m}(1+z)^{3} + (1-\Omega_{m})E^{2n},
\end{equation}
which admits no closed-form solution for $n\neq0$. Its implementation in \texttt{Kosmulator} is straightforward:
\begin{lstlisting}[style=python]
from scipy.optimize import fsolve
def f1CDM_MODEL(z: Number, p: Dict[str, float]) -> Number:
    """f1CDM implicit model: E^2 = Om(1+z)^3 + (1-Om) E^(2n)
    Solved per redshift using a scalar root finder."""
    Om = float(p["Omega_m"])
    n  = float(p["n"])
    z_arr = _asarray(z)
    out = np.empty_like(z_arr, dtype=float)

    for i, zi in enumerate(z_arr):
        def eq(E):
            return E**2 - (Om*(1.0+zi)**3 + (1.0-Om)*E**(2.0*n))

        E0 = np.sqrt(Om*(1.0+zi)**3 + (1.0-Om))  # ΛCDM seed
        try:
            sol = fsolve(eq, x0=E0, xtol=1e-8, maxfev=100)[0]
            out[i] = sol if np.isfinite(sol) and sol > 0.0 else np.nan
        except Exception:
            out[i] = np.nan

    return _scalar_or_array(out)
\end{lstlisting}
As illustrated, the overall structure of the model function remains unchanged; the user simply introduces a suitable numerical root-finding procedure. It should be noted that due to limitations of the iterative solvers, care must be taken to select a solver appropriate to the model in order to avoid numerical instabilities, and to include exception handling for robustness. 

This approach applies to all background and late-time observational probes supported by \texttt{Kosmulator}. The primary exception is CMB analysis, which requires perturbation-level equations to be implemented within \texttt{CLASS}, via a Taylor expansion ~\cite{Nesseris2013}, to compute theoretical power spectra; for such cases, we implemented an interface within \texttt{Kosmulator} to communicate directly with \texttt{CLASS} and \texttt{CLIK} for the specified model (briefly discussed in Section \ref{sec: CMB integration}). 

\subsubsection{Vectorised models and observation pipelines}

A key performance-oriented feature of \texttt{Kosmulator} is support for vectorised model and observation evaluations. In this context, vectorisation refers to evaluating both background cosmological quantities (most notably the expansion rate $E(z)$) and the corresponding observation-model predictions (e.g.\ $H(z)$, distance moduli, and growth observables) for an entire array of redshifts or data points in a single function call, using array-based arithmetic rather than explicit Python loops.

To illustrate the idea intuitively, consider a simple function such as $f(x)=3x-2$. When sketching this function by hand, one does not explicitly compute $f(x)$ for each individual value of $x$; instead, the functional form is applied simultaneously across the domain, i.e., a straight line going through $f(0)=-2$ with a gradient of $+3$. In contrast, many numerical implementations evaluate functions point-by-point, looping over each input value and storing the result in an array. While straightforward, such scalar evaluation introduces significant overhead which becomes increasingly computationally expensive when repeated many times during an MCMC sampling process.

Vectorised implementations instead apply the same mathematical operations simultaneously to all input values, delegating the computation to optimised numerical backends such as \texttt{NumPy}. This substantially reduces Python-level overhead and improves cache efficiency. In \texttt{Kosmulator}, vectorisation extends beyond the background expansion history itself. When a model is flagged as vectorised, the full observation pipeline, including distance calculations, Hubble-rate evaluations, and phenomenological growth predictions, is evaluated in an array-native manner within the \texttt{Zeus} ensamble likelihood function. As a result, a single model evaluation produces the complete vector of theoretical predictions required for a given dataset, rather than looping over individual data points.

For analytically solvable models, vectorisation is typically straightforward. For non-analytically solvable or implicitly defined models, vectorisation can still be achieved by implementing array-based numerical solvers. As an example, the implicit $f(Q)$-inspired model defined by Eq.~(\ref{eq:f1CDM_model}) can be implemented in a fully vectorised form by replacing a scalar root-finding routine with an array-based Newton--Raphson scheme. An example implementation is shown below:
\begin{lstlisting}[style=python]
def f1CDM_MODEL_vectorised(z: Number, p: Dict[str, float]) -> Number:
    """Vectorised f1CDM solver using Newton-Raphson."""
    z = np.asarray(z)
    Om, n = float(p["Omega_m"]), float(p["n"])
    
    # Seed with LCDM solution (vectorised)
    E = np.sqrt(Om * (1 + z)**3 + (1 - Om))

    # Newton-Raphson iteration on full array
    for _ in range(60):
        f = E**2 - (Om * (1 + z)**3 + (1 - Om) * E**(2.0*n))
        df = 2.0 * E - (1.0 - Om) * (2.0*n) * E**(2.0*n - 1.0)
        
        step = f / df
        E -= step
        
        if np.all(np.abs(step) < 1e-8): break
            
    return E
\end{lstlisting}
As shown, the overall structure of the model function remains unchanged; the primary difference lies in replacing a scalar solver with an array-based iterative scheme and introducing appropriate safeguards for numerical stability. This allows even implicitly defined Friedmann equations to benefit from vectorised execution. 

To ensure predictable and reproducible performance, vectorised model evaluations in \texttt{Kosmulator} are executed in a strictly single-process, single-thread configuration. Internal BLAS, OpenMP, and related numerical backends are constrained to a single thread, avoiding hidden parallelism, thread oversubscription, and non-deterministic scaling. The resulting performance gains arise primarily from reduced Python overhead rather than implicit multithreading.

Certain datasets, most notably CMB and BBN likelihoods, necessarily rely on external solvers such as \texttt{CLASS}, \texttt{CLIK}, or \texttt{AlterBBN}, and therefore remain non-vectorised in the current implementation.

The computational benefit of vectorisation is most pronounced for moderate walker counts and late-time datasets. For very large walker numbers or environments with abundant computational resources, parallel execution may become more efficient.

\subsubsection{Built-in information criteria and model comparison}

\texttt{Kosmulator} includes built-in support for standard information-criterion-based model comparison. In particular, users may designate a \emph{reference model} against which alternative or modified cosmological models are evaluated. In practice, this reference model is often chosen to be $\Lambda$CDM, reflecting its status as the current concordance cosmology, although other baseline models, such as the $w_0w_a$CDM model employed in recent DESI analyses~\cite{Abdul2025}, may also be used.

Given a specified baseline, \texttt{Kosmulator} automatically computes a set of widely adopted diagnostics for each candidate model, including the best-fit likelihood, reduced $\chi^2$, the Akaike Information Criterion (AIC)~\cite{Akaike1974}, and the Bayesian Information Criterion (BIC)~\cite{Schwarz1978}. Differences in these metrics ($\Delta$AIC and $\Delta$BIC) are evaluated relative to the reference model to provide a clear indication of relative performance while penalising unnecessary model complexity.

\texttt{Kosmulator} reports these statistics in a consistent, publication-ready format and uses them to highlight potential issues such as overfitting or lack of statistical support for additional model parameters. The intent is not to replace a full Bayesian model-comparison analysis, but rather to provide immediate, interpretable guidance during exploratory model development.

At present, the statistical analysis is limited to these standard information-criterion-based comparisons~\cite{Akaike1974,Schwarz1978}. Future development plans include the possible incorporation of additional diagnostics such as the Deviance Information Criterion (DIC)~\cite{Spiegelhalter2002}, the Watanabe--Akaike Information Criterion (WAIC)~\cite{Watanabe2010}, Leave-One-Out Cross-Validation (LOO-CV)~\cite{Vehtari2015}, Bayesian evidence and Bayes factors~\cite{Kass1995}, small-sample-corrected variants such as AICc~\cite{Hurvich1989}, the Hannan--Quinn Information Criterion (HQIC)~\cite{Hannan1979}, and focused information criteria (FIC)~\cite{Claeskens2003}, subject to feasibility and computational cost.

\subsubsection{Structure growth}
Kosmulator provides optional likelihood support for phenomenological structure-growth observables, specifically the growth rate $f(z)$ and redshift-space distortion observable $f_{\sigma_{8}}(z)$~\cite{Kazantzidis2018, Perenon2019}. 
In this approach, the growth of linear matter perturbations is modelled using the growth-index parameterisation
\begin{equation}
  f(z) \equiv \frac{d\ln D}{d\ln a} = \Omega^\gamma_m(z),
\end{equation}
where $D(a)$ is the linear growth factor, $\Omega_{m}$ is derived from the background expansion history $E(z)$, and $\gamma$ is treated as either a free phenomenological parameter or fixed to a GR-motivated value.

Predictions for $f_{\sigma_{8}}(z)$ are obtained by combining this parameterisation with the evolution of the growth factor,
\begin{equation}
  f_{\sigma_{8}}(z) = \sigma_{8}\Omega_{m}(z)^{\gamma}\exp\left[ -\int^{z}_{0} \frac{\Omega_{m}(z^{\prime})^{\gamma}}{1+z^{\prime}}dz^{\prime}\right],
\end{equation}
allowing direct comparison with observational growth measurements without requiring perturbation-level calculations. In~\cite{Sahlu2024}, the structure growth observations were added to \texttt{Kosmulator}.This framework intentionally decouples clustering inference from Boltzmann solvers, relying only on the background expansion history $E(z)$. As a result, it enables exploration of how matter clusters in deviations from General Relativity. However, a critical distinction remains: this prescription assumes scale-independent growth, effectively isolating the influence of the background geometry on structure formation. It is, therefore, a ``geometric consistency check'' rather than a full perturbation analysis. It cannot, for instance, capture the scale-dependent suppression or enhancement of power often found in screened modified gravity theories (e.g., Chameleon mechanisms in $f(R)$~\cite{Khoury2004}) without further modification. Consequently, this module should be viewed as a high-speed diagnostic tool—a means to flag potential growth tensions before committing to computationally expensive, scale-dependent perturbation solvers.

\subsubsection{CMB Integration}\label{sec: CMB integration}
While \texttt{Kosmulator} focuses on modular Pythonic implementations for background cosmology, it supports early-universe probes via a wrapper for the \texttt{CLASS} Boltzmann solver~\cite{Lesgourgues2011}. The package automatically manages the interface with the solver, handling parameter mapping and nuisance parameters for Planck 2018 likelihoods (TT, TE, EE, and lensing) transparently. This allows researchers to combine custom background models with standard CMB constraints without modifying any underlying C-code. A more detailed exploration of this integration will be presented in a forthcoming research article.

% ---------------------------------------------------------
\section{Validation and Performance Showcase}
To verify the accuracy of our likelihood implementations, the runs in Table~\ref{tab:soft_launch_results} compare \texttt{Kosmulator} against a \texttt{Cobaya} $\Lambda$CDM background only baseline\footnote{Note: \texttt{Cobaya} was not utilised for the non-linear model which includes the $n$ parameter, since the constraining of $f_{1}$CDM model is purely for illustrative purpose to show \texttt{Kosmulator's} ease for fitting non-linear models.}. We contrast the industry-standard Metropolis-Hastings sampler (via \texttt{Cobaya}) with the \texttt{Zeus} ensemble slice sampler (via \texttt{Kosmulator}). \texttt{Zeus} is chosen specifically to demonstrate the efficacy of the vectorised likelihood: because the vectorised approach is built to run on a single CPU core to maximize array efficiency, it allows us to quantify the speed recovered by avoiding standard Python loops and MPI overhead\footnote{While \texttt{EMCEE} is supported, the current vectorised likelihood implementation is optimized for the \texttt{Zeus} sampler interface. \texttt{EMCEE} can still fit non-analytical models, but typically relies on MPI parallelisation across multiple cores rather than single-core vectorisation.}.

The observational suite, consisting of \textit{Pantheon+SH0ES}, \textit{DESI DR2} BAO, and \textit{CC}, was selected to provide a robust anchor for the expansion history. \textit{Pantheon+SH0ES} provides a calibrated distance ladder necessary for constraining $H_{0}$ and the absolute magnitude $M_{abs}$. We included the \textit{DESI DR2} BAO data as it represents the current frontier of baryon acoustic oscillation measurements, offering tight constraints on the transverse and radial distance scales \cite{Abdul2025}. Finally, \textit{CC} were added because they offer a direct, model-independent measurement of $H(z)$. This is particularly useful for breaking degeneracies between, for example, $M_{abs}$, $H_{0}$, and $r_{d}$, that can be present in various cosmologies, such as the well-known $H_{0}$ and $r_{d}$ degeneracy found in fitting the $\Lambda$model directly to BAO-related datasets, such as \textit{DESI DR2}~\cite{Bernal2016, Knox2020}.

Crucially, the runtimes reported correspond to the wall-clock time required to reach statistical convergence, rather than a fixed number of steps. For \texttt{Cobaya}, we utilized the standard Gelman-Rubin ($R-1$) stopping criterion on a background-only cosmology. For \texttt{Kosmulator}, we configured the samplers with a relaxed iteration cap, but enabled dynamic termination based on the integrated autocorrelation time ($\tau$). Lastly, we added a run using the non-analytical $f_{1}$CDM model to show that the internal numerical solvers can handle models without closed-form expansion histories without a significant performance penalty.

It is worth noting that a direct comparison of iterations per second ($it/s$) between Metropolis-Hastings and Ensemble Slice Sampling is conservative. Slice sampling typically requires significantly more likelihood evaluations per ``step'' to guarantee detailed balance; however, it often produces samples with a lower autocorrelation time (higher effective sample size) than Metropolis chains. Thus, the raw speed-up shown in Table~\ref{tab:soft_launch_results} likely understates the true computational advantage: \texttt{Kosmulator} generates independent samples significantly faster (approaching an order of magnitude in wall-clock time) on a fraction of the hardware.

\subsection{Discussion of Results}
\begin{figure}[h]
    \centering
    % First Subfigure (Left)
    \begin{subfigure}[b]{0.49\textwidth}
        \centering
        % Replace with your actual filename for Plot 1
        \includegraphics[width=\textwidth]{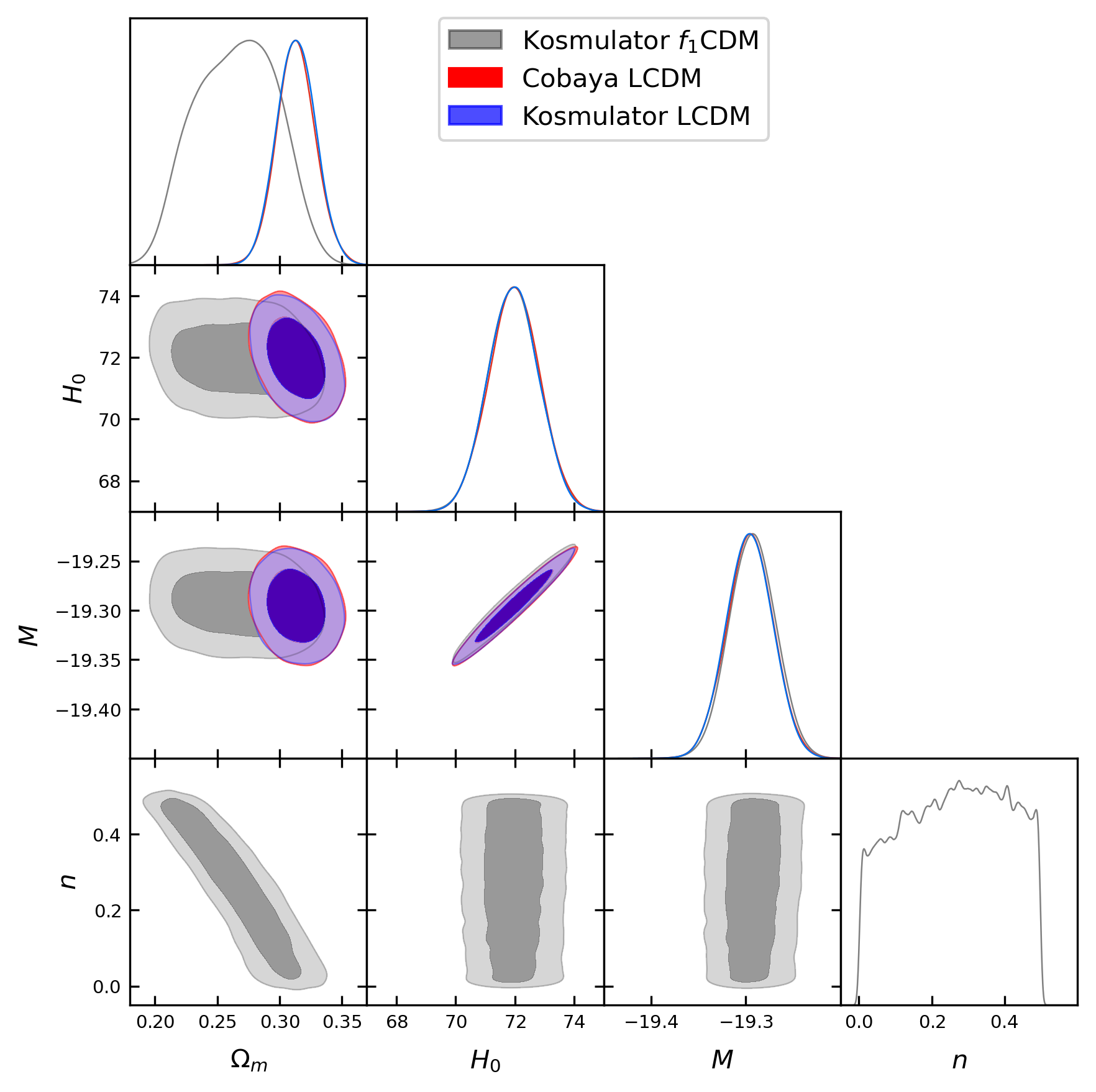}
        \caption{Pantheon$^{+}$ + SH0ES + CC}
        \label{fig:comparison_plot_cc}
    \end{subfigure}
    \hfill % Adds flexible space between the images
    % Second Subfigure (Right)
    \begin{subfigure}[b]{0.49\textwidth}
        \centering
        % Replace with your actual filename for Plot 2
        \includegraphics[width=\textwidth]{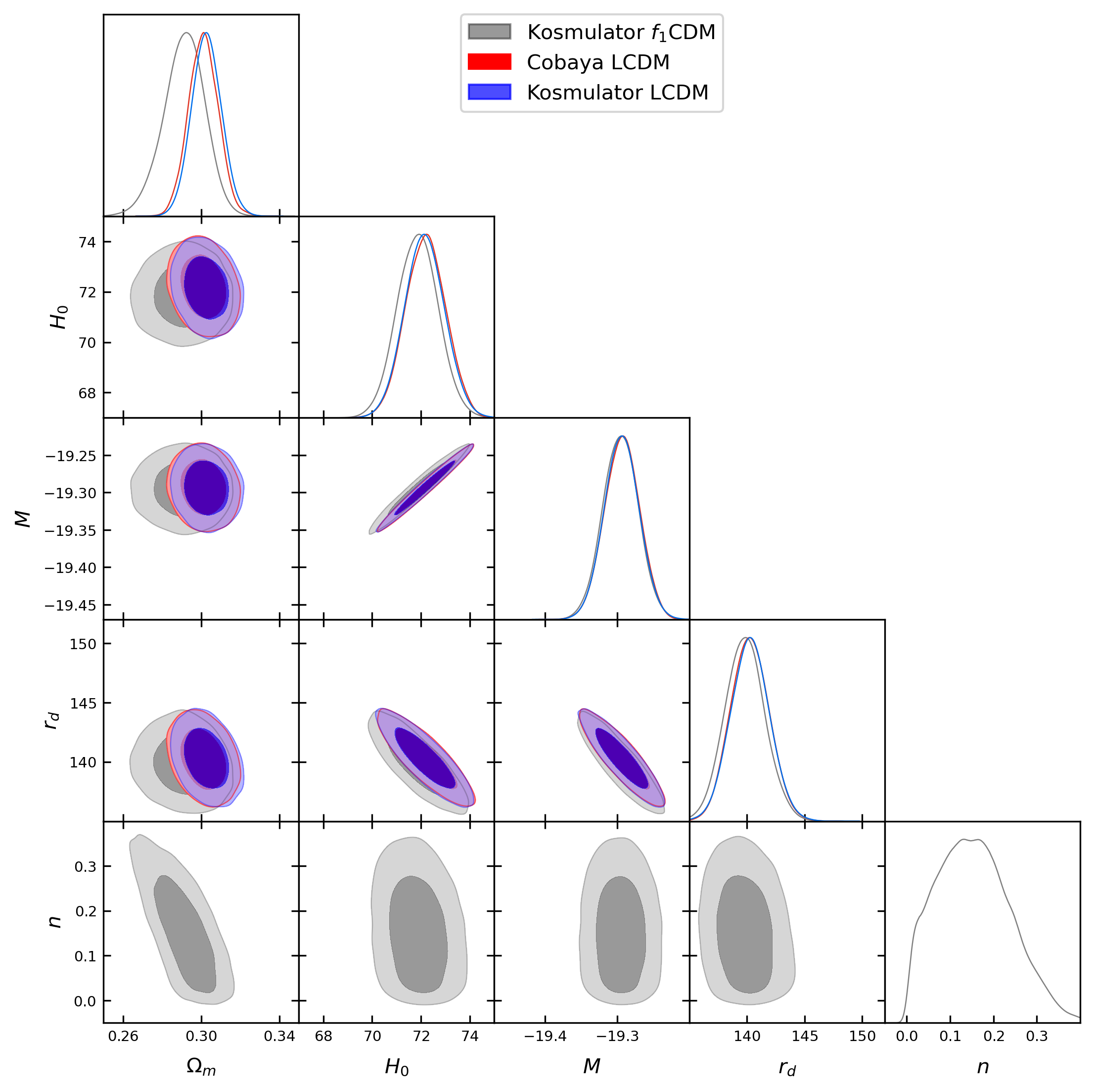}
        \caption{Pantheon$^{+}$ + SH0ES + DESI DR2 + CC}
        \label{fig:comparison_plot_desi}
    \end{subfigure}
    
    \caption{Corner plots comparing the posterior distributions obtained from \texttt{Kosmulator} (Blue/Black) and \texttt{Cobaya} (Red). \textbf{Left:} Results for the Pantheon$^{+}$ + SH0ES + CC combination. \textbf{Right:} Results including DESI DR2 BAO data, showing the tightening of constraints and the robust agreement between pipelines.}
    \label{fig:comparison_plot}
\end{figure}

The numerical validation is visualized in Figure~\ref{fig:comparison_plot}. The left panel displays the constraints from the joint \textit{Pantheon+SH0ES} and \textit{CC} likelihood combination. The contours from \texttt{Kosmulator} (Blue) and \texttt{Cobaya} (Red) are virtually indistinguishable, validating the precision of the background evolution solvers. As shown in Table~\ref{tab:soft_launch_results}, the best-fit $\chi^2$ values agree to within $\leq0.6\%$, and the derived constraints on $H_0$ and $\Omega_m$ are consistent to within $0.3\sigma$.

The performance differential is substantial. For the $\Lambda$CDM background run, \texttt{Kosmulator} achieved convergence in just \textbf{2 minutes and 9 seconds using a single CPU core}. In contrast, the \texttt{Cobaya} baseline required \textbf{9 minutes and 44 seconds distributed across 12 cores}. This represents a wall-clock speedup of $\sim 4.5\times$ while utilizing $1/12^{th}$ of the computational resources. This efficiency gain is critical for future surveys, where the volume of data will require millions of likelihood evaluations that would otherwise be computationally prohibitive.

\begin{table}[h]
\centering
\caption{Baseline consistency and performance showcase. The $f_1$CDM model introduces $n$ as an additional degree of freedom. Comparison highlights the efficiency of the single-core \texttt{Kosmulator} pipeline against the MPI-parallelised \texttt{Cobaya} baseline.}
\label{tab:soft_launch_results}
\footnotesize
\renewcommand{\arraystretch}{1.45}
\setlength{\tabcolsep}{2.6pt}

\begin{adjustbox}{max width=\linewidth}
\begin{tabular}{l l r r r r r r r}
\hline\hline
\textbf{Model} & \textbf{Pipeline} & \textbf{Sampler} & \textbf{Cores} & \textbf{Time} & \textbf{Iter.} & \textbf{It/s} & \textbf{$\chi^2$} &  \\
 & \textit{Constraints} & $H_0$ & $\Omega_m$ & $r_d$ & $M_{abs}$ & $\Omega_b h^2$ & $\Omega_{\mathrm{CDM}} h^2$ & $n$ \\
\hline

% --- DATASET 1 ---
\multicolumn{9}{c}{\textit{\textbf{Dataset: Pantheon$^{+}$+ SH0ES  + CC (Background Only)}}} \\
$\Lambda$CDM & \textbf{Cobaya}  & Metropolis & 12 & 9m 44s & 105{,}717 & 181.02 & 1486.84 &  \\
& \textit{Results} & $71.986^{+0.861}_{-0.861}$ & $^\dagger0.313^{+0.016}_{-0.016}$ & --- & $ -19.296^{+0.025}_{-0.025}$ & $0.0199^{+0.0058}_{-0.0058}$ & $0.142^{+0.010}_{-0.010}$ & --- \\
\hline

$\Lambda$CDM & \textbf{Kosmulator} & Zeus & 1 & 2m 09s & 57{,}600 & 443.88 & 1478.46 &  \\
& \textit{Results} & $71.955^{+0.856}_{-0.860}$ & $0.313^{+0.016}_{-0.016}$ & --- & $-19.296^{+0.024}_{-0.024}$ & --- & --- & --- \\
\hline

$f_{1}$CDM & \textbf{Kosmulator} & Zeus & 1 & 3m 39s & 64{,}800 & 295.89 & 1477.46 &  \\
& \textit{Results} & $71.955^{+0.868}_{-0.858}$ & $0.266^{+0.033}_{-0.036}$ & --- & $-19.293^{+0.025}_{-0.025}$ & --- & --- & $0.267^{+0.154}_{-0.168}$ \\
\hline

% --- DATASET 2 ---
\multicolumn{9}{c}{\textit{\textbf{Dataset: Pantheon$^{+}$+ SH0ES + CC + BAO}}} \\
$\Lambda$CDM & \textbf{Cobaya $r_{d}$-calibrated} & Metropolis & 12 & 10m 12s & 56{,}980 & 93.10 & 1498.69 &  \\
& \textit{Results} & $72.194^{+0.819}_{-0.819}$ & $^\dagger0.301^{+0.008}_{-0.008}$ & $140.254^{+1.669}_{-1.669}$ & $-19.293^{+0.024}_{-0.024}$ & $0.0265^{+0.0012}_{-0.0012}$ & $0.131^{+0.004}_{-0.004}$ & --- \\
\hline

$\Lambda$CDM & \textbf{Kosmulator $r_{d}$-free} & Zeus & 1 & 2m 10s & 61{,}200 & 470.77 & 1489.41 &  \\
& \textit{Results} & $72.144^{+0.821}_{-0.803}$ & $0.303^{+0.008}_{-0.008}$ & $140.274^{+1.676}_{-1.686}$ & $-19.294^{+0.024}_{-0.024}$ & --- & --- & --- \\
\hline

$f_1$CDM & \textbf{Kosmulator $r_{d}$-free} & Zeus & 1 & 6m 24s & 86{,}400 & 225.00 & 1487.42 &  \\
& \textit{Results} & $71.875^{+0.828}_{-0.841}$ & $0.292^{+0.010}_{-0.011}$ & $139.810^{+1.704}_{-1.715}$ & $-19.295^{+0.024}_{-0.024}$ & --- & --- & $0.150^{+0.092}_{-0.086}$ \\
\hline

% --- DATASET 3 ---
%\multicolumn{9}{c}{\textit{\textbf{Dataset: Pantheon$^{+}$+ SH0ES + CC + BAO+BBN\_AlterBBN (Inverse Distance Ladder) - $r_{d}$-Calibrated}}} \\
%$\Lambda$CDM & \textbf{Kosmulator} & EMCEE & 12 & 2h 16m 39 & 169{,}200 & 24.12 & 1505.20 &  \\
%& \textit{Results} & $69.404^{+0.375}_{-0.373}$ & $0.311^{+0.008}_{-0.007}$ & $148.866^{+1.136}_{-1.120}$ & $-19.374^{+0.013}_{-0.013}$ & $0.0219^{+0.0002}_{-0.0002}$ & --- & --- \\
\hline

%$f_1$CDM & \textbf{Kosmulator} & EMCEE & 12 & 4h 49m 16s & 363{,}600 & 225.00 & 1506.38 &  \\
%& \textit{Results} & $69.176^{+0.419}_{-0.437}$ & $0.310^{+0.008}_{-0.007}$ & $149.222^{+1.212}_{-1.138}$ & $-19.380^{+0.014}_{-0.014}$ & $0.0219^{+0.0002}_{-0.0002}$ &  & $0.015^{+0.023}_{-0.011}$ \\
%\hline\hline

\multicolumn{9}{l}{$^\dagger$ \textit{Derived from primary sampling of $\Omega_b h^2 + \Omega_{\mathrm{CDM}} h^2$.}}\\
\end{tabular}
\end{adjustbox}
\end{table}

The right panel of Figure~\ref{fig:comparison_plot} and the second section of Table~\ref{tab:soft_launch_results} introduce the \textit{DESI DR2} BAO data. Notably, even with the addition of the $r_d$ parameter (which \texttt{Kosmulator} samples directly as a free parameter (for this particular case), whereas standard \texttt{Cobaya} runs often treat it as a derived parameter in this configuration), the agreement remains robust with a similar time speed up. The $f_1$CDM model runs further demonstrate that expanding the parameter space to include the viscosity term $n$ incurs a negligible time penalty (between 1.5-4 minutes additional runtime for this particular model and observations), validating the solver's stability even in non-standard cosmological scenarios \footnote{It should be noted that these comparison runs utilise baseline configurations. A more comprehensive performance analysis covering high-dimensional parameter spaces will be presented in the forthcoming article.}.

\subsection{Automated Model Selection Features}

A key design philosophy of \texttt{Kosmulator} is to streamline the interpretation of results. Beyond standard parameter constraints, the pipeline natively computes model selection statistics (specifically the AIC and BIC criteria) and their relative differences ($\Delta$) between competing models.

Crucially, the software outputs an automated natural-language interpretation of these statistics based on the Jeffreys' scale (which can be found in the Statistical Analysis Tables folder). This assists inexperienced users in immediately identifying whether a new model is statistically preferred or not. Table~\ref{tab:model_selection} presents the direct output of this module for the $f_1$CDM vs $\Lambda$CDM comparison. While the $f_1$CDM model achieves a slightly lower $\chi^2$, the \texttt{Kosmulator} report correctly identifies that the additional complexity is penalized by the BIC, providing ``moderate evidence against'' the extended model model for the combined datasets. This automation removes a common manual step in the theoretical workflow, allowing researchers to assess model viability instantly.

\begin{table}[h]
\centering
\caption{Automated model selection statistics generated by \texttt{Kosmulator}. ($\Lambda$CDM) was chosen as the reference model, against which the software interprets the significance of the $\Delta$ values for the alternative ($f_1$CDM) model. Note the contrast: $\Delta$AIC ssuggests the models are statistically indistinguishable, while the stricter penalty for the extra parameter in $\Delta$BIC reveals a preference for the simpler $\Lambda$CDM model.}
\label{tab:model_selection}
\footnotesize
\renewcommand{\arraystretch}{1.3}
\setlength{\tabcolsep}{3pt}

\begin{tabular}{l c c c c c l}
\hline\hline
\textbf{Model} & $\mathbf{\chi^2_{min}}$ & \textbf{AIC} & $\mathbf{\Delta}$\textbf{AIC} & \textbf{BIC} & $\mathbf{\Delta}$\textbf{BIC} & \textbf{Automated Interpretation} \\
\hline
\multicolumn{7}{l}{\textit{Dataset: Pantheon$^{+}$ + SH0ES + CC}} \\
$\Lambda$CDM (Ref) & 1478.46 & 1484.46 & 0.0 & 1500.77 & 0.0 & --- \\
$f_1$CDM & 1477.46 & 1485.46 & +1.00 & 1507.20 & +6.43 & AIC -indistinguishable, BIC-Moderate against \\
\hline
\multicolumn{7}{l}{\textit{Dataset: Pantheon$^{+}$ + SH0ES + CC + DESI}} \\
$\Lambda$CDM (Ref) & 1489.39 & 1497.39 & 0.0 & 1519.16 & 0.0 & --- \\
$f_1$CDM & 1487.42 &  1497.42 & + 0.03 & 1524.63 & +5.47 & AIC -indistinguishable, BIC-Moderate against \\
\hline\hline
\end{tabular}
\end{table}

% ---------------------------------------------------------
\section{Discussion: The ``Fail-Fast'' Philosophy}

\texttt{Kosmulator} is intended to complement, rather than replace, high-precision frameworks like \texttt{Cobaya}. Its primary value lies in the exploratory phase of model building—a ``Fail-Fast'' prototyping stage. 

By using a pure-Python environment with vectorised likelihoods, the ``hypothesis-to-test'' cycle is reduced from weeks of C-level modifications to minutes of Python scripting. If a proposed modified gravity model cannot achieve a statistically competitive fit at the background or phenomenological growth level using \texttt{Kosmulator's} fast check, it is unlikely to justify the significant development effort required for a full perturbation-level analysis needed in \texttt{Cobaya/CLASS}. 

\subsection{Utility and Practicality}
\begin{itemize}
    \item \textbf{High-Throughput Screening:} The ability to run full MCMC chains for background-only models in minutes on standard hardware allows for the rapid elimination of non-viable theories.
    \item \textbf{Modularity:} Because the codebase avoids complex dependency chains and recompilation, it is well-suited for collaborative projects where different team members may need to inject custom phenomenological parameters or experimental datasets quickly.
    \item \textbf{Limitations as a Feature:} By explicitly focusing on late-time and background observables, \texttt{Kosmulator} maintains a small footprint. It serves as the ``scientific sieve'' that identifies high-potential candidates for subsequent, rigorous investigation using full Einstein--Boltzmann solvers.
\end{itemize}

\subsection{Future Outlook: Towards Vectorised Perturbations}
While the current version of \texttt{Kosmulator} serves primarily as a rapid screening tool, our long-term vision includes bridging the gap between exploratory prototyping and full-scale precision cosmology. A key area of interest is the potential implementation of a natively vectorised Einstein--Boltzmann solver within the framework. 

By applying the same vectorisation principles used here for background evolution to the linear perturbation equations, we aim to match the physical accuracy of established industry standards while maintaining the speed advantages of array-native execution. This would allow \texttt{Kosmulator} to handle high-dimensional, early-universe parameter spaces with the same efficiency currently seen in late-time probes. While this represents a significant computational and theoretical undertaking, it remains the ultimate goal for the framework's evolution, to provide a single, high-performance pipeline that supports the entire research cycle from initial hypothesis to final precision constraint. Whether this goal is even possible remains to be seen.

% ---------------------------------------------------------
\section{Conclusion}
In this work, we have introduced \texttt{Kosmulator}, a modular and vectorised Python framework designed to accelerate the ``hypothesis-to-test'' cycle in theoretical cosmology. As the landscape of modified gravity and dark energy models expands, the computational cost of implementing every new theory into full Einstein--Boltzmann solvers has become a significant bottleneck. \texttt{Kosmulator} addresses this by providing a lightweight, pure-Python environment that prioritises implementation speed and array-native execution.

Our validation results against the industry-standard \texttt{Cobaya} framework demonstrate that this approach requires no compromise on accuracy. By leveraging single-core vectorisation and dynamic MCMC termination, \texttt{Kosmulator} reproduced $\Lambda$CDM constraints to within $<0.3\sigma$ statistical agreement on $H_{0}$ and $\Omega_{m}$ and $\leq0.6\%$ precision while delivering a substantial increase in sampling efficiency—outperforming a massive MPI-parallelised baseline using only a fraction of the hardware resources. Furthermore, the successful implementation of the implicit $f_{1}$CDM model illustrates the framework's capability to handle non-analytical Friedmann equations, while the integrated automated model selection module (AIC/BIC) ensures that the statistical viability of such extensions is assessed immediately alongside the parameter constraints.

Ultimately, \texttt{Kosmulator} is not designed to replace high-precision perturbation codes, but to complement them. By allowing researchers to statistically rule out non-viable models in minutes rather than weeks, it focuses community resources on the most promising theoretical candidates. Looking forward, we aim to extend this vectorised philosophy to linear perturbation theory, moving towards a fully array-native Boltzmann solver. The code is open-source and available to the community at \url{https://github.com/renierht/Kosmulator}, offering a flexible testbed for the next generation of cosmological theories.

% ---------------------------------------------------------
\section*{Acknowledgements}

RH acknowledges the utilisation of \texttt{ChatGPT-5.2} and \texttt{Gemini-3} for assistance with manuscript preparation and grammatical corrections.

% ---------------------------------------------------------
\bibliographystyle{unsrt}
\bibliography{references}

@ARTICLE{Abbott2022,
       author = {{Abbott}, T.~M.~C. and {Aguena}, M. and {Alarcon}, A. and {Allam}, S. and {Alves}, O. and {Amon}, A. and {Andrade-Oliveira}, F. and {Annis}, J. and {Avila}, S. and {Bacon}, D. and {Baxter}, E. and {Bechtol}, K. and {Becker}, M.~R. and {Bernstein}, G.~M. and {Bhargava}, S. and {Birrer}, S. and {Blazek}, J. and {Brandao-Souza}, A. and {Bridle}, S.~L. and {Brooks}, D. and {Buckley-Geer}, E. and {Burke}, D.~L. and {Camacho}, H. and {Campos}, A. and {Carnero Rosell}, A. and {Carrasco Kind}, M. and {Carretero}, J. and {Castander}, F.~J. and {Cawthon}, R. and {Chang}, C. and {Chen}, A. and {Chen}, R. and {Choi}, A. and {Conselice}, C. and {Cordero}, J. and {Costanzi}, M. and {Crocce}, M. and {da Costa}, L.~N. and {da Silva Pereira}, M.~E. and {Davis}, C. and {Davis}, T.~M. and {De Vicente}, J. and {DeRose}, J. and {Desai}, S. and {Di Valentino}, E. and {Diehl}, H.~T. and {Dietrich}, J.~P. and {Dodelson}, S. and {Doel}, P. and {Doux}, C. and {Drlica-Wagner}, A. and {Eckert}, K. and {Eifler}, T.~F. and {Elsner}, F. and {Elvin-Poole}, J. and {Everett}, S. and {Evrard}, A.~E. and {Fang}, X. and {Farahi}, A. and {Fernandez}, E. and {Ferrero}, I. and {Fert{\'e}}, A. and {Fosalba}, P. and {Friedrich}, O. and {Frieman}, J. and {Garc{\'\i}a-Bellido}, J. and {Gatti}, M. and {Gaztanaga}, E. and {Gerdes}, D.~W. and {Giannantonio}, T. and {Giannini}, G. and {Gruen}, D. and {Gruendl}, R.~A. and {Gschwend}, J. and {Gutierrez}, G. and {Harrison}, I. and {Hartley}, W.~G. and {Herner}, K. and {Hinton}, S.~R. and {Hollowood}, D.~L. and {Honscheid}, K. and {Hoyle}, B. and {Huff}, E.~M. and {Huterer}, D. and {Jain}, B. and {James}, D.~J. and {Jarvis}, M. and {Jeffrey}, N. and {Jeltema}, T. and {Kovacs}, A. and {Krause}, E. and {Kron}, R. and {Kuehn}, K. and {Kuropatkin}, N. and {Lahav}, O. and {Leget}, P.-F. and {Lemos}, P. and {Liddle}, A.~R. and {Lidman}, C. and {Lima}, M. and {Lin}, H. and {MacCrann}, N. and {Maia}, M.~A.~G. and {Marshall}, J.~L. and {Martini}, P. and {McCullough}, J. and {Melchior}, P. and {Mena-Fern{\'a}ndez}, J. and {Menanteau}, F. and {Miquel}, R. and {Mohr}, J.~J. and {Morgan}, R. and {Muir}, J. and {Myles}, J. and {Nadathur}, S. and {Navarro-Alsina}, A. and {Nichol}, R.~C. and {Ogando}, R.~L.~C. and {Omori}, Y. and {Palmese}, A. and {Pandey}, S. and {Park}, Y. and {Paz-Chinch{\'o}n}, F. and {Petravick}, D. and {Pieres}, A. and {Plazas Malag{\'o}n}, A.~A. and {Porredon}, A. and {Prat}, J. and {Raveri}, M. and {Rodriguez-Monroy}, M. and {Rollins}, R.~P. and {Romer}, A.~K. and {Roodman}, A. and {Rosenfeld}, R. and {Ross}, A.~J. and {Rykoff}, E.~S. and {Samuroff}, S. and {S{\'a}nchez}, C. and {Sanchez}, E. and {Sanchez}, J. and {Sanchez Cid}, D. and {Scarpine}, V. and {Schubnell}, M. and {Scolnic}, D. and {Secco}, L.~F. and {Serrano}, S. and {Sevilla-Noarbe}, I. and {Sheldon}, E. and {Shin}, T. and {Smith}, M. and {Soares-Santos}, M. and {Suchyta}, E. and {Swanson}, M.~E.~C. and {Tabbutt}, M. and {Tarle}, G. and {Thomas}, D. and {To}, C. and {Troja}, A. and {Troxel}, M.~A. and {Tucker}, D.~L. and {Tutusaus}, I. and {Varga}, T.~N. and {Walker}, A.~R. and {Weaverdyck}, N. and {Wechsler}, R. and {Weller}, J. and {Yanny}, B. and {Yin}, B. and {Zhang}, Y. and {Zuntz}, J. and {DES Collaboration}},
        title = "{Dark Energy Survey Year 3 results: Cosmological constraints from galaxy clustering and weak lensing}",
      journal = {\prd},
         year = 2022,
        month = jan,
       volume = {105},
       number = {2},
          eid = {023520},
        pages = {023520},
          doi = {10.1103/PhysRevD.105.023520},
archivePrefix = {arXiv},
       eprint = {2105.13549},
 primaryClass = {astro-ph.CO},
       adsurl = {https://ui.adsabs.harvard.edu/abs/2022PhRvD.105b3520A}
}

@ARTICLE{Abdalla2022,
       author = {{Abdalla}, Elcio and {Abell{\'a}n}, Guillermo Franco and {Aboubrahim}, Amin and {Agnello}, Adriano and {Akarsu}, {\"O}zg{\"u}r and {Akrami}, Yashar and {Alestas}, George and {Aloni}, Daniel and {Amendola}, Luca and {Anchordoqui}, Luis A. and {Anderson}, Richard I. and {Arendse}, Nikki and {Asgari}, Marika and {Ballardini}, Mario and {Barger}, Vernon and {Basilakos}, Spyros and {Batista}, Ronaldo C. and {Battistelli}, Elia S. and {Battye}, Richard and {Benetti}, Micol and {Benisty}, David and {Berlin}, Asher and {de Bernardis}, Paolo and {Berti}, Emanuele and {Bidenko}, Bohdan and {Birrer}, Simon and {Blakeslee}, John P. and {Boddy}, Kimberly K. and {Bom}, Clecio R. and {Bonilla}, Alexander and {Borghi}, Nicola and {Bouchet}, Fran{\c{c}}ois R. and {Braglia}, Matteo and {Buchert}, Thomas and {Buckley-Geer}, Elizabeth and {Calabrese}, Erminia and {Caldwell}, Robert R. and {Camarena}, David and {Capozziello}, Salvatore and {Casertano}, Stefano and {Chen}, Geoff C.-F. and {Chluba}, Jens and {Chen}, Angela and {Chen}, Hsin-Yu and {Chudaykin}, Anton and {Cicoli}, Michele and {Copi}, Craig J. and {Courbin}, Fred and {Cyr-Racine}, Francis-Yan and {Czerny}, Bo{\.z}ena and {Dainotti}, Maria and {D'Amico}, Guido and {Davis}, Anne-Christine and {de Cruz P{\'e}rez}, Javier and {de Haro}, Jaume and {Delabrouille}, Jacques and {Denton}, Peter B. and {Dhawan}, Suhail and {Dienes}, Keith R. and {Di Valentino}, Eleonora and {Du}, Pu and {Eckert}, Dominique and {Escamilla-Rivera}, Celia and {Fert{\'e}}, Agn{\`e}s and {Finelli}, Fabio and {Fosalba}, Pablo and {Freedman}, Wendy L. and {Frusciante}, Noemi and {Gazta{\~n}aga}, Enrique and {Giar{\`e}}, William and {Giusarma}, Elena and {G{\'o}mez-Valent}, Adri{\`a} and {Handley}, Will and {Harrison}, Ian and {Hart}, Luke and {Hazra}, Dhiraj Kumar and {Heavens}, Alan and {Heinesen}, Asta and {Hildebrandt}, Hendrik and {Hill}, J. Colin and {Hogg}, Natalie B. and {Holz}, Daniel E. and {Hooper}, Deanna C. and {Hosseininejad}, Nikoo and {Huterer}, Dragan and {Ishak}, Mustapha and {Ivanov}, Mikhail M. and {Jaffe}, Andrew H. and {Jang}, In Sung and {Jedamzik}, Karsten and {Jimenez}, Raul and {Joseph}, Melissa and {Joudaki}, Shahab and {Kamionkowski}, Marc and {Karwal}, Tanvi and {Kazantzidis}, Lavrentios and {Keeley}, Ryan E. and {Klasen}, Michael and {Komatsu}, Eiichiro and {Koopmans}, L{\'e}on V.~E. and {Kumar}, Suresh and {Lamagna}, Luca and {Lazkoz}, Ruth and {Lee}, Chung-Chi and {Lesgourgues}, Julien and {Levi Said}, Jackson and {Lewis}, Tiffany R. and {L'Huillier}, Benjamin and {Lucca}, Matteo and {Maartens}, Roy and {Macri}, Lucas M. and {Marfatia}, Danny and {Marra}, Valerio and {Martins}, Carlos J.~A.~P. and {Masi}, Silvia and {Matarrese}, Sabino and {Mazumdar}, Arindam and {Melchiorri}, Alessandro and {Mena}, Olga and {Mersini-Houghton}, Laura and {Mertens}, James and {Milakovi{\'c}}, Dinko and {Minami}, Yuto and {Miranda}, Vivian and {Moreno-Pulido}, Cristian and {Moresco}, Michele and {Mota}, David F. and {Mottola}, Emil and {Mozzon}, Simone and {Muir}, Jessica and {Mukherjee}, Ankan and {Mukherjee}, Suvodip and {Naselsky}, Pavel and {Nath}, Pran and {Nesseris}, Savvas and {Niedermann}, Florian and {Notari}, Alessio and {Nunes}, Rafael C. and {{\'O} Colg{\'a}in}, Eoin and {Owens}, Kayla A. and {{\"O}z{\"u}lker}, Emre and {Pace}, Francesco and {Paliathanasis}, Andronikos and {Palmese}, Antonella and {Pan}, Supriya and {Paoletti}, Daniela and {Perez Bergliaffa}, Santiago E. and {Perivolaropoulos}, Leandros and {Pesce}, Dominic W. and {Pettorino}, Valeria and {Philcox}, Oliver H.~E. and {Pogosian}, Levon and {Poulin}, Vivian and {Poulot}, Gaspard and {Raveri}, Marco and {Reid}, Mark J. and {Renzi}, Fabrizio and {Riess}, Adam G. and {Sabla}, Vivian I. and {Salucci}, Paolo and {Salzano}, Vincenzo and {Saridakis}, Emmanuel N. and {Sathyaprakash}, Bangalore S. and {Schmaltz}, Martin and {Sch{\"o}neberg}, Nils and {Scolnic}, Dan and {Sen}, Anjan A. and {Sehgal}, Neelima and {Shafieloo}, Arman and {Sheikh-Jabbari}, M.~M. and {Silk}, Joseph and {Silvestri}, Alessandra and {Skara}, Foteini and {Sloth}, Martin S. and {Soares-Santos}, Marcelle and {Sol{\`a} Peracaula}, Joan and {Songsheng}, Yu-Yang and {Soriano}, Jorge F. and {Staicova}, Denitsa and {Starkman}, Glenn D. and {Szapudi}, Istv{\'a}n and {Teixeira}, Elsa M. and {Thomas}, Brooks and {Treu}, Tommaso and {Trott}, Emery and {van de Bruck}, Carsten and {Vazquez}, J. Alberto and {Verde}, Licia and {Visinelli}, Luca and {Wang}, Deng and {Wang}, Jian-Min and {Wang}, Shao-Jiang and {Watkins}, Richard and {Watson}, Scott and {Webb}, John K. and {Weiner}, Neal and {Weltman}, Amanda and {Witte}, Samuel J. and {Wojtak}, Rados{\l}aw and {Yadav}, Anil Kumar},
        title = "{Cosmology intertwined: A review of the particle physics, astrophysics, and cosmology associated with the cosmological tensions and anomalies}",
      journal = {Journal of High Energy Astrophysics},
         year = 2022,
        month = jun,
       volume = {34},
        pages = {49-211},
          doi = {10.1016/j.jheap.2022.04.002},
archivePrefix = {arXiv},
       eprint = {2203.06142},
 primaryClass = {astro-ph.CO},
       adsurl = {https://ui.adsabs.harvard.edu/abs/2022JHEAp..34...49A}
}

@ARTICLE{Abdul2025,
       author = {{Abdul Karim}, M. and {Aguilar}, J. and {Ahlen}, S. and {Alam}, S. and {Allen}, L. and {Prieto}, C. Allende and {Alves}, O. and {Anand}, A. and {Andrade}, U. and {Armengaud}, E. and {Aviles}, A. and {Bailey}, S. and {Baltay}, C. and {Bansal}, P. and {Bault}, A. and {Behera}, J. and {BenZvi}, S. and {Bianchi}, D. and {Blake}, C. and {Brieden}, S. and {Brodzeller}, A. and {Brooks}, D. and {Buckley-Geer}, E. and {Burtin}, E. and {Calderon}, R. and {Canning}, R. and {Rosell}, A. Carnero and {Carrilho}, P. and {Casas}, L. and {Castander}, F.~J. and {Charles}, M. and {Chaussidon}, E. and {Chaves-Montero}, J. and {Chebat}, D. and {Chen}, X. and {Claybaugh}, T. and {Cole}, S. and {Cooper}, A.~P. and {Cuceu}, A. and {Dawson}, K.~S. and {de la Macorra}, A. and {de Mattia}, A. and {Deiosso}, N. and {Della Costa}, J. and {Demina}, R. and {Dey}, A. and {Dey}, B. and {Ding}, Z. and {Doel}, P. and {Edelstein}, J. and {Eisenstein}, D.~J. and {Elbers}, W. and {Fagrelius}, P. and {Fanning}, K. and {Fern{\'a}ndez-Garc{\'\i}a}, E. and {Ferraro}, S. and {Font-Ribera}, A. and {Forero-Romero}, J.~E. and {Frenk}, C.~S. and {Garcia-Quintero}, C. and {Garrison}, L.~H. and {Gazta{\~n}aga}, E. and {Gil-Mar{\'\i}n}, H. and {Gontcho A Gontcho}, S. and {Gonzalez}, D. and {Gonzalez-Morales}, A.~X. and {Gordon}, C. and {Green}, D. and {Gutierrez}, G. and {Guy}, J. and {Hadzhiyska}, B. and {Hahn}, C. and {He}, S. and {Herbold}, M. and {Herrera-Alcantar}, H.~K. and {Ho}, M.-F. and {Honscheid}, K. and {Howlett}, C. and {Huterer}, D. and {Ishak}, M. and {Juneau}, S. and {Kamble}, N.~V. and {Kara{\c{c}}ayl{\i}}, N.~G. and {Kehoe}, R. and {Kent}, S. and {Kim}, A.~G. and {Kirkby}, D. and {Kisner}, T. and {Koposov}, S.~E. and {Kremin}, A. and {Krolewski}, A. and {Lahav}, O. and {Lamman}, C. and {Landriau}, M. and {Lang}, D. and {Lasker}, J. and {Le Goff}, J.~M. and {Le Guillou}, L. and {Leauthaud}, A. and {Levi}, M.~E. and {Li}, Q. and {Li}, T.~S. and {Lodha}, K. and {Lokken}, M. and {Lozano-Rodr{\'\i}guez}, F. and {Magneville}, C. and {Manera}, M. and {Martini}, P. and {Matthewson}, W.~L. and {Meisner}, A. and {Mena-Fern{\'a}ndez}, J. and {Menegas}, A. and {Mergulh{\~a}o}, T. and {Miquel}, R. and {Moustakas}, J. and {Mu{\~n}oz-Guti{\'e}rrez}, A. and {Mu{\~n}oz-Santos}, D. and {Myers}, A.~D. and {Nadathur}, S. and {Naidoo}, K. and {Napolitano}, L. and {Newman}, J.~A. and {Niz}, G. and {Noriega}, H.~E. and {Paillas}, E. and {Palanque-Delabrouille}, N. and {Pan}, J. and {Peacock}, J.~A. and {Pellejero Ibanez}, M. and {Percival}, W.~J. and {P{\'e}rez-Fern{\'a}ndez}, A. and {P{\'e}rez-R{\`a}fols}, I. and {Pieri}, M.~M. and {Poppett}, C. and {Prada}, F. and {Rabinowitz}, D. and {Raichoor}, A. and {Ram{\'\i}rez-P{\'e}rez}, C. and {Rashkovetskyi}, M. and {Ravoux}, C. and {Rich}, J. and {Rocher}, A. and {Rockosi}, C. and {Rohlf}, J. and {Rom{\'a}n-Herrera}, J.~O. and {Ross}, A.~J. and {Rossi}, G. and {Ruggeri}, R. and {Ruhlmann-Kleider}, V. and {Samushia}, L. and {Sanchez}, E. and {Sanders}, N. and {Schlegel}, D. and {Schubnell}, M. and {Seo}, H. and {Shafieloo}, A. and {Sharples}, R. and {Silber}, J. and {Sinigaglia}, F. and {Sprayberry}, D. and {Tan}, T. and {Tarl{\'e}}, G. and {Taylor}, P. and {Turner}, W. and {Ure{\~n}a-L{\'o}pez}, L.~A. and {Vaisakh}, R. and {Valdes}, F. and {Valogiannis}, G. and {Vargas-Maga{\~n}a}, M. and {Verde}, L. and {Walther}, M. and {Weaver}, B.~A. and {Weinberg}, D.~H. and {White}, M. and {Wolfson}, M. and {Y{\`e}che}, C. and {Yu}, J. and {Zaborowski}, E.~A. and {Zarrouk}, P. and {Zhai}, Z. and {Zhang}, H. and {Zhao}, C. and {Zhao}, G.~B. and {Zhou}, R. and {Zou}, H. and {DESI Collaboration}},
        title = "{DESI DR2 results. II. Measurements of baryon acoustic oscillations and cosmological constraints}",
      journal = {\prd},
         year = 2025,
        month = oct,
       volume = {112},
       number = {8},
          eid = {083515},
        pages = {083515},
          doi = {10.1103/tr6y-kpc6},
archivePrefix = {arXiv},
       eprint = {2503.14738},
 primaryClass = {astro-ph.CO},
       adsurl = {https://ui.adsabs.harvard.edu/abs/2025PhRvD.112h3515A}
}

@ARTICLE{Adame2025,
       author = {{Adame}, A.~G. and {Aguilar}, J. and {Ahlen}, S. and {Alam}, S. and {Alexander}, D.~M. and {Alvarez}, M. and {Alves}, O. and {Anand}, A. and {Andrade}, U. and {Armengaud}, E. and {Avila}, S. and {Aviles}, A. and {Awan}, H. and {Bahr-Kalus}, B. and {Bailey}, S. and {Baltay}, C. and {Bault}, A. and {Behera}, J. and {BenZvi}, S. and {Bera}, A. and {Beutler}, F. and {Bianchi}, D. and {Blake}, C. and {Blum}, R. and {Brieden}, S. and {Brodzeller}, A. and {Brooks}, D. and {Buckley-Geer}, E. and {Burtin}, E. and {Calderon}, R. and {Canning}, R. and {Carnero Rosell}, A. and {Cereskaite}, R. and {Cervantes-Cota}, J.~L. and {Chabanier}, S. and {Chaussidon}, E. and {Chaves-Montero}, J. and {Chen}, S. and {Chen}, X. and {Claybaugh}, T. and {Cole}, S. and {Cuceu}, A. and {Davis}, T.~M. and {Dawson}, K. and {de la Macorra}, A. and {de Mattia}, A. and {Deiosso}, N. and {Dey}, A. and {Dey}, B. and {Ding}, Z. and {Doel}, P. and {Edelstein}, J. and {Eftekharzadeh}, S. and {Eisenstein}, D.~J. and {Elliott}, A. and {Fagrelius}, P. and {Fanning}, K. and {Ferraro}, S. and {Ereza}, J. and {Findlay}, N. and {Flaugher}, B. and {Font-Ribera}, A. and {Forero-S{\'a}nchez}, D. and {Forero-Romero}, J.~E. and {Frenk}, C.~S. and {Garcia-Quintero}, C. and {Gazta{\~n}aga}, E. and {Gil-Mar{\'\i}n}, H. and {Gontcho a Gontcho}, S. and {Gonzalez-Morales}, A.~X. and {Gonzalez-Perez}, V. and {Gordon}, C. and {Green}, D. and {Gruen}, D. and {Gsponer}, R. and {Gutierrez}, G. and {Guy}, J. and {Hadzhiyska}, B. and {Hahn}, C. and {Hanif}, M.~M.~S. and {Herrera-Alcantar}, H.~K. and {Honscheid}, K. and {Howlett}, C. and {Huterer}, D. and {Ir{\v{s}}i{\v{c}}}, V. and {Ishak}, M. and {Juneau}, S. and {Kara{\c{c}}ayl{\i}}, N.~G. and {Kehoe}, R. and {Kent}, S. and {Kirkby}, D. and {Kremin}, A. and {Krolewski}, A. and {Lai}, Y. and {Lan}, T.-W. and {Landriau}, M. and {Lang}, D. and {Lasker}, J. and {Le Goff}, J.~M. and {Le Guillou}, L. and {Leauthaud}, A. and {Levi}, M.~E. and {Li}, T.~S. and {Linder}, E. and {Lodha}, K. and {Magneville}, C. and {Manera}, M. and {Margala}, D. and {Martini}, P. and {Maus}, M. and {McDonald}, P. and {Medina-Varela}, L. and {Meisner}, A. and {Mena-Fern{\'a}ndez}, J. and {Miquel}, R. and {Moon}, J. and {Moore}, S. and {Moustakas}, J. and {Mueller}, E. and {Mu{\~n}oz-Guti{\'e}rrez}, A. and {Myers}, A.~D. and {Nadathur}, S. and {Napolitano}, L. and {Neveux}, R. and {Newman}, J.~A. and {Nguyen}, N.~M. and {Nie}, J. and {Niz}, G. and {Noriega}, H.~E. and {Padmanabhan}, N. and {Paillas}, E. and {Palanque-Delabrouille}, N. and {Pan}, J. and {Penmetsa}, S. and {Percival}, W.~J. and {Pieri}, M.~M. and {Pinon}, M. and {Poppett}, C. and {Porredon}, A. and {Prada}, F. and {P{\'e}rez-Fern{\'a}ndez}, A. and {P{\'e}rez-R{\`a}fols}, I. and {Rabinowitz}, D. and {Raichoor}, A. and {Ram{\'\i}rez-P{\'e}rez}, C. and {Ramirez-Solano}, S. and {Rashkovetskyi}, M. and {Ravoux}, C. and {Rezaie}, M. and {Rich}, J. and {Rocher}, A. and {Rockosi}, C. and {Roe}, N.~A. and {Rosado-Marin}, A. and {Ross}, A.~J. and {Rossi}, G. and {Ruggeri}, R. and {Ruhlmann-Kleider}, V. and {Samushia}, L. and {Sanchez}, E. and {Saulder}, C. and {Schlafly}, E.~F. and {Schlegel}, D. and {Schubnell}, M. and {Seo}, H. and {Shafieloo}, A. and {Sharples}, R. and {Silber}, J. and {Slosar}, A. and {Smith}, A. and {Sprayberry}, D. and {Tan}, T. and {Tarl{\'e}}, G. and {Taylor}, P. and {Trusov}, S. and {Ure{\~n}a-L{\'o}pez}, L.~A. and {Vaisakh}, R. and {Valcin}, D. and {Valdes}, F. and {Vargas-Maga{\~n}a}, M. and {Verde}, L. and {Walther}, M. and {Wang}, B. and {Wang}, M.~S. and {Weaver}, B.~A. and {Weaverdyck}, N. and {Wechsler}, R.~H. and {Weinberg}, D.~H. and {White}, M. and {Yu}, J. and {Yu}, Y. and {Yuan}, S. and {Y{\`e}che}, C. and {Zaborowski}, E.~A. and {Zarrouk}, P. and {Zhang}, H. and {Zhao}, C. and {Zhao}, R. and {Zhou}, R. and {Zhuang}, T.},
        title = "{DESI 2024 VI: cosmological constraints from the measurements of baryon acoustic oscillations}",
      journal = {\jcap},
         year = 2025,
        month = feb,
       volume = {2025},
       number = {2},
          eid = {021},
        pages = {021},
          doi = {10.1088/1475-7516/2025/02/021},
archivePrefix = {arXiv},
       eprint = {2404.03002},
 primaryClass = {astro-ph.CO},
       adsurl = {https://ui.adsabs.harvard.edu/abs/2025JCAP...02..021A}
}

@ARTICLE{Akaike1974,
       author = {{Akaike}, H.},
        title = "{A New Look at the Statistical Model Identification}",
      journal = {IEEE Transactions on Automatic Control},
         year = 1974,
        month = jan,
       volume = {19},
        pages = {716-723},
          doi = {10.1109/TAC.1974.1100705},
       adsurl = {https://ui.adsabs.harvard.edu/abs/1974ITAC...19..716A}
}

@ARTICLE{Asgari2021,
       author = {{Asgari}, Marika and {Lin}, Chieh-An and {Joachimi}, Benjamin and {Giblin}, Benjamin and {Heymans}, Catherine and {Hildebrandt}, Hendrik and {Kannawadi}, Arun and {St{\"o}lzner}, Benjamin and {Tr{\"o}ster}, Tilman and {van den Busch}, Jan Luca and {Wright}, Angus H. and {Bilicki}, Maciej and {Blake}, Chris and {de Jong}, Jelte and {Dvornik}, Andrej and {Erben}, Thomas and {Getman}, Fedor and {Hoekstra}, Henk and {K{\"o}hlinger}, Fabian and {Kuijken}, Konrad and {Miller}, Lance and {Radovich}, Mario and {Schneider}, Peter and {Shan}, HuanYuan and {Valentijn}, Edwin},
        title = "{KiDS-1000 cosmology: Cosmic shear constraints and comparison between two point statistics}",
      journal = {\aap},
         year = 2021,
        month = jan,
       volume = {645},
          eid = {A104},
        pages = {A104},
          doi = {10.1051/0004-6361/202039070},
archivePrefix = {arXiv},
       eprint = {2007.15633},
 primaryClass = {astro-ph.CO},
       adsurl = {https://ui.adsabs.harvard.edu/abs/2021A&A...645A.104A}
}

@ARTICLE{Beltran2017,
       author = {{Beltran Jimenez}, Jose and {Heisenberg}, Lavinia and {Koivisto}, Tomi},
        title = "{Coincident General Relativity}",
      journal = {arXiv e-prints},
         year = 2017,
        month = oct,
          eid = {arXiv:1710.03116},
        pages = {arXiv:1710.03116},
          doi = {10.48550/arXiv.1710.03116},
archivePrefix = {arXiv},
       eprint = {1710.03116},
 primaryClass = {gr-qc},
       adsurl = {https://ui.adsabs.harvard.edu/abs/2017arXiv171003116B}
}

@ARTICLE{Bengochea2009,
       author = {{Bengochea}, Gabriel R. and {Ferraro}, Rafael},
        title = "{Dark torsion as the cosmic speed-up}",
      journal = {\prd},
         year = 2009,
        month = jun,
       volume = {79},
       number = {12},
          eid = {124019},
        pages = {124019},
          doi = {10.1103/PhysRevD.79.124019},
archivePrefix = {arXiv},
       eprint = {0812.1205},
 primaryClass = {astro-ph},
       adsurl = {https://ui.adsabs.harvard.edu/abs/2009PhRvD..79l4019B}
}

@ARTICLE{Bernal2016,
       author = {{Bernal}, Jos{\'e} Luis and {Verde}, Licia and {Riess}, Adam G.},
        title = "{The trouble with H$_{0}$}",
      journal = {\jcap},
         year = 2016,
        month = oct,
       volume = {2016},
       number = {10},
          eid = {019},
        pages = {019},
          doi = {10.1088/1475-7516/2016/10/019},
archivePrefix = {arXiv},
       eprint = {1607.05617},
 primaryClass = {astro-ph.CO},
       adsurl = {https://ui.adsabs.harvard.edu/abs/2016JCAP...10..019B}
}

@ARTICLE{Brinckmann2019,
       author = {{Brinckmann}, Thejs and {Lesgourgues}, Julien},
        title = "{MontePython 3: Boosted MCMC sampler and other features}",
      journal = {Physics of the Dark Universe},
         year = 2019,
        month = mar,
       volume = {24},
          eid = {100260},
        pages = {100260},
          doi = {10.1016/j.dark.2018.100260},
archivePrefix = {arXiv},
       eprint = {1804.07261},
 primaryClass = {astro-ph.CO},
       adsurl = {https://ui.adsabs.harvard.edu/abs/2019PDU....24..260B}
}

@ARTICLE{Brout2022,
       author = {{Brout}, Dillon and {Scolnic}, Dan and {Popovic}, Brodie and {Riess}, Adam G. and {Carr}, Anthony and {Zuntz}, Joe and {Kessler}, Rick and {Davis}, Tamara M. and {Hinton}, Samuel and {Jones}, David and {Kenworthy}, W. D'Arcy and {Peterson}, Erik R. and {Said}, Khaled and {Taylor}, Georgie and {Ali}, Noor and {Armstrong}, Patrick and {Charvu}, Pranav and {Dwomoh}, Arianna and {Meldorf}, Cole and {Palmese}, Antonella and {Qu}, Helen and {Rose}, Benjamin M. and {Sanchez}, Bruno and {Stubbs}, Christopher W. and {Vincenzi}, Maria and {Wood}, Charlotte M. and {Brown}, Peter J. and {Chen}, Rebecca and {Chambers}, Ken and {Coulter}, David A. and {Dai}, Mi and {Dimitriadis}, Georgios and {Filippenko}, Alexei V. and {Foley}, Ryan J. and {Jha}, Saurabh W. and {Kelsey}, Lisa and {Kirshner}, Robert P. and {M{\"o}ller}, Anais and {Muir}, Jessie and {Nadathur}, Seshadri and {Pan}, Yen-Chen and {Rest}, Armin and {Rojas-Bravo}, Cesar and {Sako}, Masao and {Siebert}, Matthew R. and {Smith}, Mat and {Stahl}, Benjamin E. and {Wiseman}, Phil},
        title = "{The Pantheon+ Analysis: Cosmological Constraints}",
      journal = {\apj},
         year = 2022,
        month = oct,
       volume = {938},
       number = {2},
          eid = {110},
        pages = {110},
          doi = {10.3847/1538-4357/ac8e04},
archivePrefix = {arXiv},
       eprint = {2202.04077},
 primaryClass = {astro-ph.CO},
       adsurl = {https://ui.adsabs.harvard.edu/abs/2022ApJ...938..110B}
}

@article{Claeskens2003,
author = {Gerda Claeskens and Nils Lid Hjort},
title = {The Focused Information Criterion},
journal = {Journal of the American Statistical Association},
volume = {98},
number = {464},
pages = {900--916},
year = {2003},
publisher = {Taylor \& Francis},
doi = {10.1198/016214503000000819},
URL = {https://doi.org/10.1198/016214503000000819},
eprint = {https://doi.org/10.1198/016214503000000819}
}

@ARTICLE{Conley2011,
       author = {{Conley}, A. and {Guy}, J. and {Sullivan}, M. and {Regnault}, N. and {Astier}, P. and {Balland}, C. and {Basa}, S. and {Carlberg}, R.~G. and {Fouchez}, D. and {Hardin}, D. and {Hook}, I.~M. and {Howell}, D.~A. and {Pain}, R. and {Palanque-Delabrouille}, N. and {Perrett}, K.~M. and {Pritchet}, C.~J. and {Rich}, J. and {Ruhlmann-Kleider}, V. and {Balam}, D. and {Baumont}, S. and {Ellis}, R.~S. and {Fabbro}, S. and {Fakhouri}, H.~K. and {Fourmanoit}, N. and {Gonz{\'a}lez-Gait{\'a}n}, S. and {Graham}, M.~L. and {Hudson}, M.~J. and {Hsiao}, E. and {Kronborg}, T. and {Lidman}, C. and {Mourao}, A.~M. and {Neill}, J.~D. and {Perlmutter}, S. and {Ripoche}, P. and {Suzuki}, N. and {Walker}, E.~S.},
        title = "{Supernova Constraints and Systematic Uncertainties from the First Three Years of the Supernova Legacy Survey}",
      journal = {\apjs},
         year = 2011,
        month = jan,
       volume = {192},
       number = {1},
          eid = {1},
        pages = {1},
          doi = {10.1088/0067-0049/192/1/1},
archivePrefix = {arXiv},
       eprint = {1104.1443},
 primaryClass = {astro-ph.CO},
       adsurl = {https://ui.adsabs.harvard.edu/abs/2011ApJS..192....1C}
}

@ARTICLE{Cooke2014,
       author = {{Cooke}, Ryan J. and {Pettini}, Max and {Jorgenson}, Regina A. and {Murphy}, Michael T. and {Steidel}, Charles C.},
        title = "{Precision Measures of the Primordial Abundance of Deuterium}",
      journal = {\apj},
         year = 2014,
        month = jan,
       volume = {781},
       number = {1},
          eid = {31},
        pages = {31},
          doi = {10.1088/0004-637X/781/1/31},
archivePrefix = {arXiv},
       eprint = {1308.3240},
 primaryClass = {astro-ph.CO},
       adsurl = {https://ui.adsabs.harvard.edu/abs/2014ApJ...781...31C}
}

@ARTICLE{DESY52024,
       author = {{DES Collaboration} and {Abbott}, T.~M.~C. and {Acevedo}, M. and {Aguena}, M. and {Alarcon}, A. and {Allam}, S. and {Alves}, O. and {Amon}, A. and {Andrade-Oliveira}, F. and {Annis}, J. and {Armstrong}, P. and {Asorey}, J. and {Avila}, S. and {Bacon}, D. and {Bassett}, B.~A. and {Bechtol}, K. and {Bernardinelli}, P.~H. and {Bernstein}, G.~M. and {Bertin}, E. and {Blazek}, J. and {Bocquet}, S. and {Brooks}, D. and {Brout}, D. and {Buckley-Geer}, E. and {Burke}, D.~L. and {Camacho}, H. and {Camilleri}, R. and {Campos}, A. and {Carnero Rosell}, A. and {Carollo}, D. and {Carr}, A. and {Carretero}, J. and {Castander}, F.~J. and {Cawthon}, R. and {Chang}, C. and {Chen}, R. and {Choi}, A. and {Conselice}, C. and {Costanzi}, M. and {da Costa}, L.~N. and {Crocce}, M. and {Davis}, T.~M. and {DePoy}, D.~L. and {Desai}, S. and {Diehl}, H.~T. and {Dixon}, M. and {Dodelson}, S. and {Doel}, P. and {Doux}, C. and {Drlica-Wagner}, A. and {Elvin-Poole}, J. and {Everett}, S. and {Ferrero}, I. and {Fert{\'e}}, A. and {Flaugher}, B. and {Foley}, R.~J. and {Fosalba}, P. and {Friedel}, D. and {Frieman}, J. and {Frohmaier}, C. and {Galbany}, L. and {Garc{\'\i}a-Bellido}, J. and {Gatti}, M. and {Gaztanaga}, E. and {Giannini}, G. and {Glazebrook}, K. and {Graur}, O. and {Gruen}, D. and {Gruendl}, R.~A. and {Gutierrez}, G. and {Hartley}, W.~G. and {Herner}, K. and {Hinton}, S.~R. and {Hollowood}, D.~L. and {Honscheid}, K. and {Huterer}, D. and {Jain}, B. and {James}, D.~J. and {Jeffrey}, N. and {Kasai}, E. and {Kelsey}, L. and {Kent}, S. and {Kessler}, R. and {Kim}, A.~G. and {Kirshner}, R.~P. and {Kovacs}, E. and {Kuehn}, K. and {Lahav}, O. and {Lee}, J. and {Lee}, S. and {Lewis}, G.~F. and {Li}, T.~S. and {Lidman}, C. and {Lin}, H. and {Malik}, U. and {Marshall}, J.~L. and {Martini}, P. and {Mena-Fern{\'a}ndez}, J. and {Menanteau}, F. and {Miquel}, R. and {Mohr}, J.~J. and {Mould}, J. and {Muir}, J. and {M{\"o}ller}, A. and {Neilsen}, E. and {Nichol}, R.~C. and {Nugent}, P. and {Ogando}, R.~L.~C. and {Palmese}, A. and {Pan}, Y.-C. and {Paterno}, M. and {Percival}, W.~J. and {Pereira}, M.~E.~S. and {Pieres}, A. and {Malag{\'o}n}, A.~A. Plazas and {Popovic}, B. and {Porredon}, A. and {Prat}, J. and {Qu}, H. and {Raveri}, M. and {Rodr{\'\i}guez-Monroy}, M. and {Romer}, A.~K. and {Roodman}, A. and {Rose}, B. and {Sako}, M. and {Sanchez}, E. and {Sanchez Cid}, D. and {Schubnell}, M. and {Scolnic}, D. and {Sevilla-Noarbe}, I. and {Shah}, P. and {Smith}, J. Allyn. and {Smith}, M. and {Soares-Santos}, M. and {Suchyta}, E. and {Sullivan}, M. and {Suntzeff}, N. and {Swanson}, M.~E.~C. and {S{\'a}nchez}, B.~O. and {Tarle}, G. and {Taylor}, G. and {Thomas}, D. and {To}, C. and {Toy}, M. and {Troxel}, M.~A. and {Tucker}, B.~E. and {Tucker}, D.~L. and {Uddin}, S.~A. and {Vincenzi}, M. and {Walker}, A.~R. and {Weaverdyck}, N. and {Wechsler}, R.~H. and {Weller}, J. and {Wester}, W. and {Wiseman}, P. and {Yamamoto}, M. and {Yuan}, F. and {Zhang}, B. and {Zhang}, Y.},
        title = "{The Dark Energy Survey: Cosmology Results with {\ensuremath{\sim}}1500 New High-redshift Type Ia Supernovae Using the Full 5 yr Data Set}",
      journal = {\apjl},
         year = 2024,
        month = sep,
       volume = {973},
       number = {1},
          eid = {L14},
        pages = {L14},
          doi = {10.3847/2041-8213/ad6f9f},
archivePrefix = {arXiv},
       eprint = {2401.02929},
 primaryClass = {astro-ph.CO},
       adsurl = {https://ui.adsabs.harvard.edu/abs/2024ApJ...973L..14D}
}

@ARTICLE{Valentino2021,
       author = {{Valentino}, Eleonora and {Mena}, Olga and {Pan}, Supriya and {Visinelli}, Luca and {Yang}, Weiqiang and {Melchiorri}, Alessandro and {Mota}, David F. and {Riess}, Adam G. and {Silk}, Joseph},
        title = "{In the realm of the Hubble tension-a review of solutions}",
      journal = {Classical and Quantum Gravity},
         year = 2021,
        month = jul,
       volume = {38},
       number = {15},
          eid = {153001},
        pages = {153001},
          doi = {10.1088/1361-6382/ac086d},
archivePrefix = {arXiv},
       eprint = {2103.01183},
 primaryClass = {astro-ph.CO},
       adsurl = {https://ui.adsabs.harvard.edu/abs/2021CQGra..38o3001D}
}

@ARTICLE{Foreman2013,
       author = {{Foreman-Mackey}, Daniel and {Hogg}, David W. and {Lang}, Dustin and {Goodman}, Jonathan},
        title = "{emcee: The MCMC Hammer}",
      journal = {\pasp},
         year = 2013,
        month = mar,
       volume = {125},
       number = {925},
        pages = {306},
          doi = {10.1086/670067},
archivePrefix = {arXiv},
       eprint = {1202.3665},
 primaryClass = {astro-ph.IM},
       adsurl = {https://ui.adsabs.harvard.edu/abs/2013PASP..125..306F}
}

@article{Hannan1979,
 ISSN = {00359246},
 URL = {http://www.jstor.org/stable/2985032},
 author = {E. J. Hannan and B. G. Quinn},
 journal = {Journal of the Royal Statistical Society. Series B (Methodological)},
 number = {2},
 pages = {190--195},
 publisher = {[Royal Statistical Society, Oxford University Press]},
 title = {The Determination of the Order of an Autoregression},
 urldate = {2026-01-19},
 volume = {41},
 year = {1979}
}

@ARTICLE{Hicken2009,
       author = {{Hicken}, Malcolm and {Challis}, Peter and {Jha}, Saurabh and {Kirshner}, Robert P. and {Matheson}, Tom and {Modjaz}, Maryam and {Rest}, Armin and {Wood-Vasey}, W. Michael and {Bakos}, Gaspar and {Barton}, Elizabeth J. and {Berlind}, Perry and {Bragg}, Ann and {Brice{\~n}o}, Cesar and {Brown}, Warren R. and {Caldwell}, Nelson and {Calkins}, Mike and {Cho}, Richard and {Ciupik}, Larry and {Contreras}, Maria and {Dendy}, Kristi-Concannon and {Dosaj}, Anil and {Durham}, Nick and {Eriksen}, Kris and {Esquerdo}, Gil and {Everett}, Mark and {Falco}, Emilio and {Fernandez}, Jose and {Gaba}, Alejandro and {Garnavich}, Peter and {Graves}, Genevieve and {Green}, Paul and {Groner}, Ted and {Hergenrother}, Carl and {Holman}, Matthew J. and {Hradecky}, Vit and {Huchra}, John and {Hutchison}, Bob and {Jerius}, Diab and {Jordan}, Andres and {Kilgard}, Roy and {Krauss}, Miriam and {Luhman}, Kevin and {Macri}, Lucas and {Marrone}, Daniel and {McDowell}, Jonathan and {McIntosh}, Daniel and {McNamara}, Brian and {Megeath}, Tom and {Mochejska}, Barbara and {Munoz}, Diego and {Muzerolle}, James and {Naranjo}, Orlando and {Narayan}, Gautham and {Pahre}, Michael and {Peters}, Wayne and {Peterson}, Dawn and {Rines}, Ken and {Ripman}, Ben and {Roussanova}, Anna and {Schild}, Rudolph and {Sicilia-Aguilar}, Aurora and {Sokoloski}, Jennifer and {Smalley}, Kyle and {Smith}, Andy and {Spahr}, Tim and {Stanek}, K.~Z. and {Barmby}, Pauline and {Blondin}, St{\'e}phane and {Stubbs}, Christopher W. and {Szentgyorgyi}, Andrew and {Torres}, Manuel A.~P. and {Vaz}, Amili and {Vikhlinin}, Alexey and {Wang}, Zhong and {Westover}, Mike and {Woods}, Deborah and {Zhao}, Ping},
        title = "{CfA3: 185 Type Ia Supernova Light Curves from the CfA}",
      journal = {\apj},
         year = 2009,
        month = jul,
       volume = {700},
       number = {1},
        pages = {331-357},
          doi = {10.1088/0004-637X/700/1/331},
archivePrefix = {arXiv},
       eprint = {0901.4787},
 primaryClass = {astro-ph.CO},
       adsurl = {https://ui.adsabs.harvard.edu/abs/2009ApJ...700..331H}
}

@ARTICLE{Hough2020,
       author = {{Hough}, R.~T. and {Abebe}, A. and {Ferreira}, S.~E.~S.},
        title = "{Viability tests of f(R)-gravity models with Supernovae Type 1A data}",
      journal = {European Physical Journal C},
         year = 2020,
        month = aug,
       volume = {80},
       number = {8},
          eid = {787},
        pages = {787},
          doi = {10.1140/epjc/s10052-020-8342-7},
archivePrefix = {arXiv},
       eprint = {1911.05983},
 primaryClass = {gr-qc},
       adsurl = {https://ui.adsabs.harvard.edu/abs/2020EPJC...80..787H}
}

@ARTICLE{Hu2007,
       author = {{Hu}, Wayne and {Sawicki}, Ignacy},
        title = "{Models of f(R) cosmic acceleration that evade solar system tests}",
      journal = {\prd},
         year = 2007,
        month = sep,
       volume = {76},
       number = {6},
          eid = {064004},
        pages = {064004},
          doi = {10.1103/PhysRevD.76.064004},
archivePrefix = {arXiv},
       eprint = {0705.1158},
 primaryClass = {astro-ph},
       adsurl = {https://ui.adsabs.harvard.edu/abs/2007PhRvD..76f4004H}
}

@article{HURVICH1989,
    author = {Hurvich, Clifford M. and Tsai, Chih-ling},
    title = {Regression and time series model selection in small samples},
    journal = {Biometrika},
    volume = {76},
    number = {2},
    pages = {297-307},
    year = {1989},
    month = {06},
    issn = {0006-3444},
    doi = {10.1093/biomet/76.2.297},
    url = {https://doi.org/10.1093/biomet/76.2.297},
    eprint = {https://academic.oup.com/biomet/article-pdf/76/2/297/737009/76-2-297.pdf},
}

@ARTICLE{Karamanis2021,
       author = {{Karamanis}, Minas and {Beutler}, Florian and {Peacock}, John A.},
        title = "{zeus: a PYTHON implementation of ensemble slice sampling for efficient Bayesian parameter inference}",
      journal = {\mnras},
         year = 2021,
        month = dec,
       volume = {508},
       number = {3},
        pages = {3589-3603},
          doi = {10.1093/mnras/stab2867},
archivePrefix = {arXiv},
       eprint = {2105.03468},
 primaryClass = {astro-ph.IM},
       adsurl = {https://ui.adsabs.harvard.edu/abs/2021MNRAS.508.3589K}
}

@article{Kass1995,
author = {Robert E. Kass and Adrian E. Raftery},
title = {Bayes Factors},
journal = {Journal of the American Statistical Association},
volume = {90},
number = {430},
pages = {773--795},
year = {1995},
publisher = {Taylor \& Francis},
doi = {10.1080/01621459.1995.10476572},
URL = {https://www.tandfonline.com/doi/abs/10.1080/01621459.1995.10476572},
eprint = {https://www.tandfonline.com/doi/pdf/10.1080/01621459.1995.10476572}

}

@ARTICLE{Kazantzidis2018,
       author = {{Kazantzidis}, Lavrentios and {Perivolaropoulos}, Leandros},
        title = "{Evolution of the f {\ensuremath{\sigma}}$_{8}$ tension with the Planck15 /{\ensuremath{\Lambda}} CDM determination and implications for modified gravity theories}",
      journal = {\prd},
         year = 2018,
        month = may,
       volume = {97},
       number = {10},
          eid = {103503},
        pages = {103503},
          doi = {10.1103/PhysRevD.97.103503},
archivePrefix = {arXiv},
       eprint = {1803.01337},
 primaryClass = {astro-ph.CO},
       adsurl = {https://ui.adsabs.harvard.edu/abs/2018PhRvD..97j3503K}
}

@ARTICLE{Khoury2004,
       author = {{Khoury}, Justin and {Weltman}, Amanda},
        title = "{Chameleon cosmology}",
      journal = {\prd},
         year = 2004,
        month = feb,
       volume = {69},
       number = {4},
          eid = {044026},
        pages = {044026},
          doi = {10.1103/PhysRevD.69.044026},
archivePrefix = {arXiv},
       eprint = {astro-ph/0309411},
 primaryClass = {astro-ph},
       adsurl = {https://ui.adsabs.harvard.edu/abs/2004PhRvD..69d4026K}
}

@ARTICLE{Knox2020,
       author = {{Knox}, L. and {Millea}, M.},
        title = "{Hubble constant hunter's guide}",
      journal = {\prd},
         year = 2020,
        month = feb,
       volume = {101},
       number = {4},
          eid = {043533},
        pages = {043533},
          doi = {10.1103/PhysRevD.101.043533},
archivePrefix = {arXiv},
       eprint = {1908.03663},
 primaryClass = {astro-ph.CO},
       adsurl = {https://ui.adsabs.harvard.edu/abs/2020PhRvD.101d3533K}
}

@ARTICLE{Lesgourgues2011,
       author = {{Lesgourgues}, Julien},
        title = "{The Cosmic Linear Anisotropy Solving System (CLASS) I: Overview}",
      journal = {arXiv e-prints},
         year = 2011,
        month = apr,
          eid = {arXiv:1104.2932},
        pages = {arXiv:1104.2932},
          doi = {10.48550/arXiv.1104.2932},
archivePrefix = {arXiv},
       eprint = {1104.2932},
 primaryClass = {astro-ph.IM},
       adsurl = {https://ui.adsabs.harvard.edu/abs/2011arXiv1104.2932L}
}

@software{Lewis2011,
       author = {{Lewis}, Antony and {Challinor}, Anthony},
        title = "{CAMB: Code for Anisotropies in the Microwave Background}",
 howpublished = {Astrophysics Source Code Library, record ascl:1102.026},
         year = 2011,
        month = feb,
          eid = {ascl:1102.026},
archivePrefix = {ascl},
       eprint = {1102.026},
       adsurl = {https://ui.adsabs.harvard.edu/abs/2011ascl.soft02026L}
}

@ARTICLE{Lodha2025,
       author = {{Lodha}, K. and {Shafieloo}, A. and {Calderon}, R. and {Linder}, E. and {Sohn}, W. and {Cervantes-Cota}, J.~L. and {de Mattia}, A. and {Garc{\'\i}a-Bellido}, J. and {Ishak}, M. and {Matthewson}, W. and {Aguilar}, J. and {Ahlen}, S. and {Brooks}, D. and {Claybaugh}, T. and {de la Macorra}, A. and {Dey}, A. and {Dey}, B. and {Doel}, P. and {Forero-Romero}, J.~E. and {Gazta{\~n}aga}, E. and {Gontcho}, S. Gontcho A. and {Howlett}, C. and {Juneau}, S. and {Kent}, S. and {Kisner}, T. and {Lambert}, A. and {Landriau}, M. and {Le Guillou}, L. and {Martini}, P. and {Meisner}, A. and {Miquel}, R. and {Moustakas}, J. and {Newman}, J.~A. and {Niz}, G. and {Palanque-Delabrouille}, N. and {Percival}, W.~J. and {Poppett}, C. and {Prada}, F. and {Rossi}, G. and {Ruhlmann-Kleider}, V. and {Sanchez}, E. and {Schlafly}, E.~F. and {Schlegel}, D. and {Schubnell}, M. and {Seo}, H. and {Sprayberry}, D. and {Tarl{\'e}}, G. and {Weaver}, B.~A. and {Zou}, H. and {DESI Collaboration}},
        title = "{DESI 2024: Constraints on physics-focused aspects of dark energy using DESI DR1 BAO data}",
      journal = {\prd},
         year = 2025,
        month = jan,
       volume = {111},
       number = {2},
          eid = {023532},
        pages = {023532},
          doi = {10.1103/PhysRevD.111.023532},
archivePrefix = {arXiv},
       eprint = {2405.13588},
 primaryClass = {astro-ph.CO},
       adsurl = {https://ui.adsabs.harvard.edu/abs/2025PhRvD.111b3532L}
}

@ARTICLE{Loubser2025a,
       author = {{Loubser}, S. Ilani and {Alabi}, Adebusola B. and {Hilton}, Matt and {Ma}, Yin-Zhe and {Tang}, Xin and {Hatamkhani}, Narges and {Cress}, Catherine and {Skelton}, Rosalind E. and {Nkosi}, S. Andile},
        title = "{An independent estimate of H(z) at z = 0.5 from the stellar ages of brightest cluster galaxies}",
      journal = {\mnras},
         year = 2025,
        month = jul,
       volume = {540},
       number = {4},
        pages = {3135-3149},
          doi = {10.1093/mnras/staf915},
archivePrefix = {arXiv},
       eprint = {2506.03836},
 primaryClass = {astro-ph.CO},
       adsurl = {https://ui.adsabs.harvard.edu/abs/2025MNRAS.540.3135L}
}

@ARTICLE{Loubser2025b,
       author = {{Loubser}, S. Ilani},
        title = "{Measuring the expansion history of the Universe with DESI cosmic chronometers}",
      journal = {\mnras},
         year = 2025,
        month = dec,
       volume = {544},
       number = {4},
        pages = {3064-3075},
          doi = {10.1093/mnras/staf1939},
archivePrefix = {arXiv},
       eprint = {2511.02730},
 primaryClass = {astro-ph.CO},
       adsurl = {https://ui.adsabs.harvard.edu/abs/2025MNRAS.544.3064L}
}

@ARTICLE{Moresco2020,
       author = {{Moresco}, Michele and {Jimenez}, Raul and {Verde}, Licia and {Cimatti}, Andrea and {Pozzetti}, Lucia},
        title = "{Setting the Stage for Cosmic Chronometers. II. Impact of Stellar Population Synthesis Models Systematics and Full Covariance Matrix}",
      journal = {\apj},
         year = 2020,
        month = jul,
       volume = {898},
       number = {1},
          eid = {82},
        pages = {82},
          doi = {10.3847/1538-4357/ab9eb0},
archivePrefix = {arXiv},
       eprint = {2003.07362},
 primaryClass = {astro-ph.GA},
       adsurl = {https://ui.adsabs.harvard.edu/abs/2020ApJ...898...82M}
}

@ARTICLE{Neill2009,
       author = {{Neill}, James D. and {Sullivan}, Mark and {Howell}, D. Andrew and {Conley}, Alex and {Seibert}, Mark and {Martin}, D. Christopher and {Barlow}, Tom A. and {Foster}, Karl and {Friedman}, Peter G. and {Morrissey}, Patrick and {Neff}, Susan G. and {Schiminovich}, David and {Wyder}, Ted K. and {Bianchi}, Luciana and {Donas}, Jos{\'e} and {Heckman}, Timothy M. and {Lee}, Young-Wook and {Madore}, Barry F. and {Milliard}, Bruno and {Rich}, R. Michael and {Szalay}, Alex S.},
        title = "{The Local Hosts of Type Ia Supernovae}",
      journal = {\apj},
         year = 2009,
        month = dec,
       volume = {707},
       number = {2},
        pages = {1449-1465},
          doi = {10.1088/0004-637X/707/2/1449},
archivePrefix = {arXiv},
       eprint = {0911.0690},
 primaryClass = {astro-ph.CO},
       adsurl = {https://ui.adsabs.harvard.edu/abs/2009ApJ...707.1449N}
}

\end{document}